\documentclass[
 reprint,
 amsmath,amssymb,
 aps,
prb,
]{revtex4-2}

\usepackage{graphicx}% Include figure files
\usepackage{dcolumn}% Align table columns on decimal point
\usepackage{bm}% bold math
\usepackage{xcolor}
\newcommand{\mb}{}
\usepackage{soul}

\usepackage{mhchem}
\usepackage{amsmath}
\usepackage{layouts}
\begin{document}
%\showthe\columnwidth
\preprint{APS/123-QED}

\title{Atomically Thin Metallenes at the Edge}

\author{Kameyab Raza Abidi}
\affiliation{Nanoscience Center, Department of Physics, University of Jyv\"askyl\"a, Finland.}
\author{Mohammad Bagheri} %https://orcid.org/0000-0002-3501-2779  
\affiliation{Nanoscience Center, Department of Physics, University of Jyv\"askyl\"a, Finland.}
\author{Sukhbir Singh}
\affiliation{Nanoscience Center, Department of Physics, University of Jyv\"askyl\"a, Finland.}
\author{Pekka Koskinen}
\affiliation{Nanoscience Center, Department of Physics, University of Jyv\"askyl\"a, Finland.}
\email{pekka.j.koskinen@jyu.fi}
 
\date{\today}

\begin{abstract}
Atomically thin metallenes are a new family of materials representing the ultimate limit of a thin free-electron gas for novel applications. Although metallene research has gained traction, limited attention has been paid to the properties of their ubiquitous edges. Here, we use density-functional theory simulations to investigate various edges of Mg, Cu, Y, Au, and Pb metallenes with hexagonal and buckled honeycomb lattices. Investigating relaxations, energies, stresses, and electronic structures at the edge, we find that some properties have clear trends while others are sensitive to both element and lattice type. Given that edge properties are fundamental to metallene stability and interactions in lateral heterostructures, their detailed understanding will help guide the development of metallene synthesis and applications.
\end{abstract}

\maketitle

%\tableofcontents

\section{Introduction}
Metallenes are a family of two-dimensional (2D) materials made of nonlayered bulk metals \cite{yang2015glitter,nevalaita2018atlas,prabhu2021metallenes,cao2022ultrathin, shahzad2024recent, Xie2023}. Their metallic bonding and nonlocalized electronic structure make them unique among 2D materials and attractive for catalytic, sensoric, biomedic, electronic, energy storage, and conversion applications \cite{Lu2023_bio, shahzad2024recent, Xie2023,yaoda2020}. Compared to covalent van der Waals materials with sample sizes in the microns \cite{tung2009high, goldene_2024}, metallenes are more delicate, with sample sizes measured in nanometers.

%At the same time, the non-directional metallic bonding and the concurrent ductility makes them also unstable against 2D clustering, making them challenging to synthesize.

In such small scales, finite-size properties are of central importance. Like most interesting phenomena happen at the surfaces of 3D bulk, so much of the same happens at the edges of 2D bulk. All atoms at 2D are surface atoms, and edge atoms are still lower-coordinated and often more reactive. Edges are central in synthesis \cite{Antikainen2017}, which may begin with metal atom decoration at the edges of covalent 2D materials \cite{liu2023novel}. Edge stability governs chemical and mechanical stability \cite{shenoy2008edge,EdgeStressInducedSpontaneous20212,Koskinen2021}. The low coordination at the edges can dominate catalytic reactions \cite{prabhu2021metallenes}. Edges are also indirectly responsible for the properties of the interfaces that stabilize metallene patches inside the pores of covalent templates \cite{Antikainen2017,Nevalaita2019,liu2023novel}. 

So far, metallene edges have been studied only in cursory and passing. Using very narrow ribbons without structural relaxations, edge energies of hexagonal and square lattices have been investigated across the metallic elements in the periodic table \cite{nevalaita2018beyond}. The stabilizing role of edges has been studied in the context of metallene patches inside covalent pores \cite{koskinen2015plenty,Nevalaita2019}. Edges have also been investigated as part of small planar gas-phase clusters \cite{koskinen2007liquid, koskinen2006density}. In these studies, the focus has been on nanostructures and finite-size quantum effects, not on edges themselves. Consequently, the detailed properties of metallene edges remain largely unknown.

Therefore, in this article, we use density-functional theory (DFT) simulations to investigate the edge relaxations, energies, stresses, and electronic structures of four different edges made of five representative metallenes Mg, Cu, Y, Au, and Pb. We find that the edge energies ($0.05-0.5$~eV/\AA), can be understood in terms of a non-linear bond-cutting model. The edge stresses ($0.04-1.18$ eV/\AA) are all tensile and small compared to ones in 2D covalent materials. Some properties behave systematically, others highly depend on element and edge type. Understanding these properties will help guide experiments to improve experimental designs for more stable and versatile metallene samples. 

\section{Edge Structures and Methods}

\begin{figure}[b!]
    \centering
    \includegraphics[width=\linewidth]{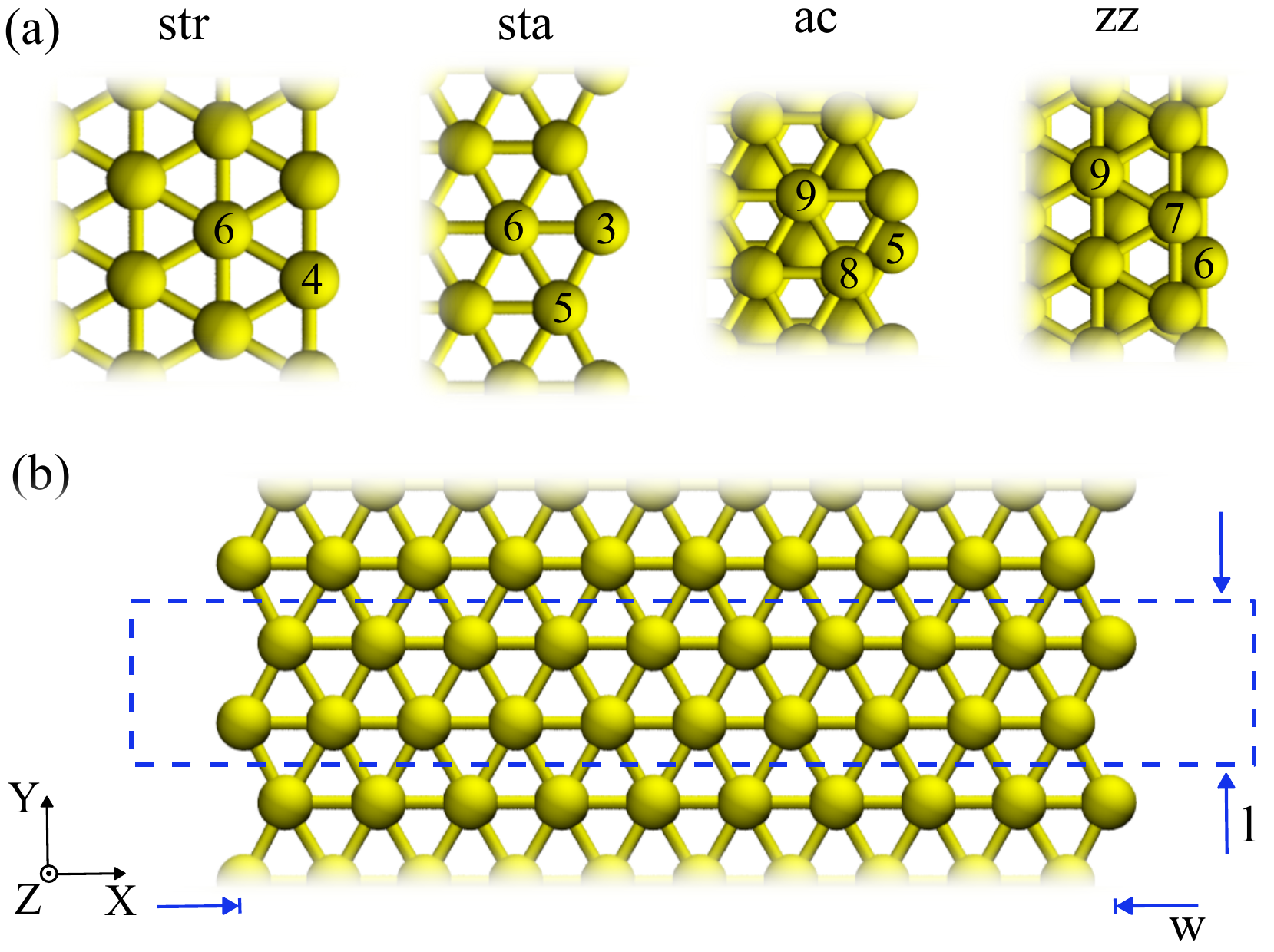}
    \caption{Studied systems and the simulation setting. a) Four edge types of hexagonal straight (str), hexagonal staggered (sta), buckled honeycomb armchair (ac), and buckled honeycomb zigzag (zz). Shown are also coordination numbers. b) The edges are modeled as a part of finite-width ribbons in a simulation cell periodic in the $y$-direction (dashed box).}
    \label{fig:systems}
\end{figure}

We investigated four different edge types: straight (str) and staggered (sta) edges of flat hexagonal lattices and zigzag (zz) and armchair (ac) edges of buckled honeycomb lattices (Fig.~\ref{fig:systems}a). We chose hexagonal and buckled honeycomb lattices for their previously identified energetic and dynamic stability among different lattices \cite{gentle}. Moreover, for the five metallenes, we chose Mg, Cu, Y, Au, and Pb because they are representative of different groups with distinct electronic configurations, and previous studies have assigned different energetic, elastic, and dynamic stability properties to them \cite{nevalaita2018atlas,nevalaita2018beyond,Nevalaita2019,gentle}. These $20$ unique edges give sufficiently representative sampling to provide comprehensive insight into the edge properties of atomically thin metallenes.

The edges were modeled using DFT implemented in the QuantumATK program suite \cite{smidstrup2019quantumatk}. The Perdew-Burke-Ernzerhow exchange-correlation functional \cite{perdew1996generalized} was used together with a local \emph{medium} basis, PseudoDojo pseudopotentials \cite{van2018pseudodojo}, and $1\times 1 \times 12$ $k$-point sampling. The Cu(str) edge was calculated with $1\times 1 \times 18$ $k$-point sampling for numerical reasons. The edges were modeled as belonging to nanoribbons of infinite length in the $y$-direction (Fig.~\ref{fig:systems}b). The non-periodic directions had a $20$~\AA\ vacuum to avoid spurious interactions between the ribbons. While the ribbon widths $w$ and the atoms in the cell $N$ were finite, we focused on the edges at the \emph{semi-infinite} limit of large $w$. Therefore, the cell lengths $l=l_0$ were derived from the 2D bulk metallene bond lengths. Under this setting, the ribbons were relaxed using the LBFGS algorithm to forces below $1$~meV/\AA\ \cite{liu1989limited,bitzek2006structural}.

The relaxed energy per unit length of such a ribbon can be expressed as
\begin{equation}
    E/l=N\varepsilon_{2D}/l + 2\lambda + \frac{1}{2}Yw\epsilon^2 + 2\tau \epsilon,
    \label{eq:edensity}
\end{equation}
where $\varepsilon_{2D}$ is 2D bulk cohesion energy, $\lambda$ is the edge energy, $Y$ is the 2D Young's modulus, $\tau$ is the edge stress, and $\epsilon=(l-l_0)/l_0$ is the axial strain. The edge energy ($\lambda>0$) is the per-unit-length cost of creating an edge into 2D bulk. The edge stress tends to either shrink ($\tau>0$, tensile) or lengthen ($\tau<0$, compressive) the edge with respect to the bulk length $l_0$. We can unambiguously assign the energies and stresses for our edges because all our ribbons have mirror (or roto-translation) symmetry, with a total edge length of $2l$ in the unit cell. For narrow ribbons, the edge energies and stresses are strongly affected by finite-size effects \cite{koskinen2008self,huang2009quantum}. However, our focus here is not on nanoscale ribbons or their width-dependent properties but on the properties of the edges of semi-infinite bulk. 

\begin{figure}
    \centering
    \includegraphics[width=1\linewidth]{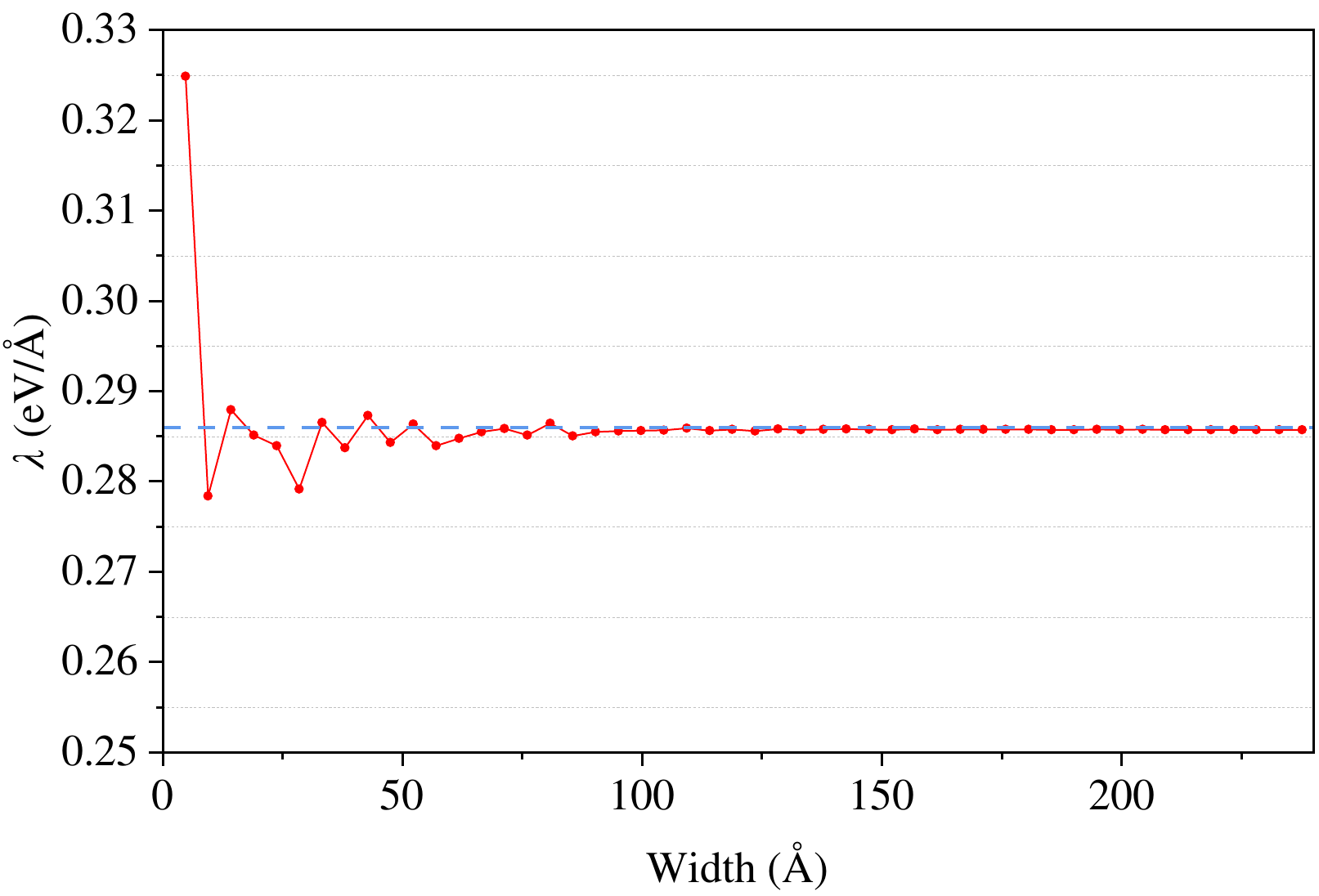}
    \caption{Fitting the edge energy. The edge energy $\lambda(w)=(E(w)-N\varepsilon_{2D})/l_0$ for the Au(str) edge (red line). The blue line shows the energy $\lambda$ of a converged, semi-infinite edge.}
    \label{fig:fitting}
\end{figure}

Despite the simple expression (\ref{eq:edensity}), robust edge energy and stress calculations need utmost care. Strictly, the edge energy from (\ref{eq:edensity}) depends on $w$ and is well-defined only at the limit $w\rightarrow \infty$; yet calculations are conducted at finite $w$. Revised periodic boundary conditions could describe edges with an extended bulk \cite{koskinen2010efficient,kit2011revised}, and Green's functions could describe edges with a semi-infinite bulk \cite{green1981, green1982, green_edge}, but such techniques are rarely available. Here, to obtain $\lambda$ reliably, we calculated ribbons with several widths at $\epsilon=0$, and fit $\varepsilon_{2D}$ and $\lambda$ to the expression $E(w)/l_0=N(w)\varepsilon_{2D}/l_0 + 2\lambda$ (Fig.~\ref{fig:fitting}) for the widest half of the calculated ribbons. In principle, $\varepsilon_{2D}$ should equal 2D bulk cohesion; in practice, the fit became more robust by adopting $\varepsilon_{2D}$ as a fit parameter. (Taking $\varepsilon_{2D}$ straight from a separate 2D bulk calculation would bring small errors that make $\lambda$ dependent on $w$ at the limit of large $N$. \mb{Still, the approach is merely a necessary technical trick without real limitations \cite{boettger1994nonconvergence, fiorentini1996extracting, nevalaita2018beyond}; the differences in the fitted and separately calculated values of 2D bulk cohesion are measured in millielectronvolts.)}

As derivatives of edge energies, the edge stresses are even more tricky to calculate. They must be calculated by straining the ribbon, which inevitably strains the bulk of the ribbon and makes bulk energies dominate when $w$ increases. Even the smallest inaccuracy in cell length $l_0$ makes the term $\tfrac{1}{2}Yw\epsilon^2$ dominant over the edge term $2\tau \epsilon$, making the determination of $\tau$ impractical for a single ribbon width. Therefore, we calculate ribbons with $32-50$ different widths $w=2.4\ldots 312$~\AA\ and $11$ different strains $\epsilon=-0.015\ldots 0.01$ and fit $E_0$, $Y$, and $\tau$ to minimize the expression
\begin{equation}
    \sum_{i,j} \left( \frac{E(w_i,\epsilon_j)}{l_0}-\left[\frac{E_0}{l_0} + \frac{1}{2}Yw_i(\epsilon_j-\epsilon')^2 + 2\tau (\epsilon_j-\epsilon')\right]\right)^2,
\end{equation}
where $\varepsilon' (\approx 0)$ is a tiny correction due to inevitable numerical inaccuracy in $l_0$; see Table S1 Electronic Supplemental Information (ESI) for the range of ribbon widths used in the fit. Using ribbons with different $w$ and $\epsilon$ simultaneously yields a fit for $\tau$ that is robust and gives a faithful estimate for the stress of a semi-infinite bulk edge. \mb{We assessed the reliability of each fit by inspecting the strain curves for all widths separately and ensuring that the obtained parameters reliably reproduce the data and have the correct physical meaning.}

Allowing for axial relaxation, the ribbon energy is minimized at $\epsilon=-2\tau/(Yw)$. Stability requirement $E>N\varepsilon_{2D}$ then implies the necessary condition $w>\tau^2/Y\lambda$, which is easily met by the typical values of $Y$ and the expected values for $\lambda$ and $\tau$. 

\section{Results}

\subsection{Edge energies}
\begin{figure}
    \centering
    \includegraphics[width=\linewidth]{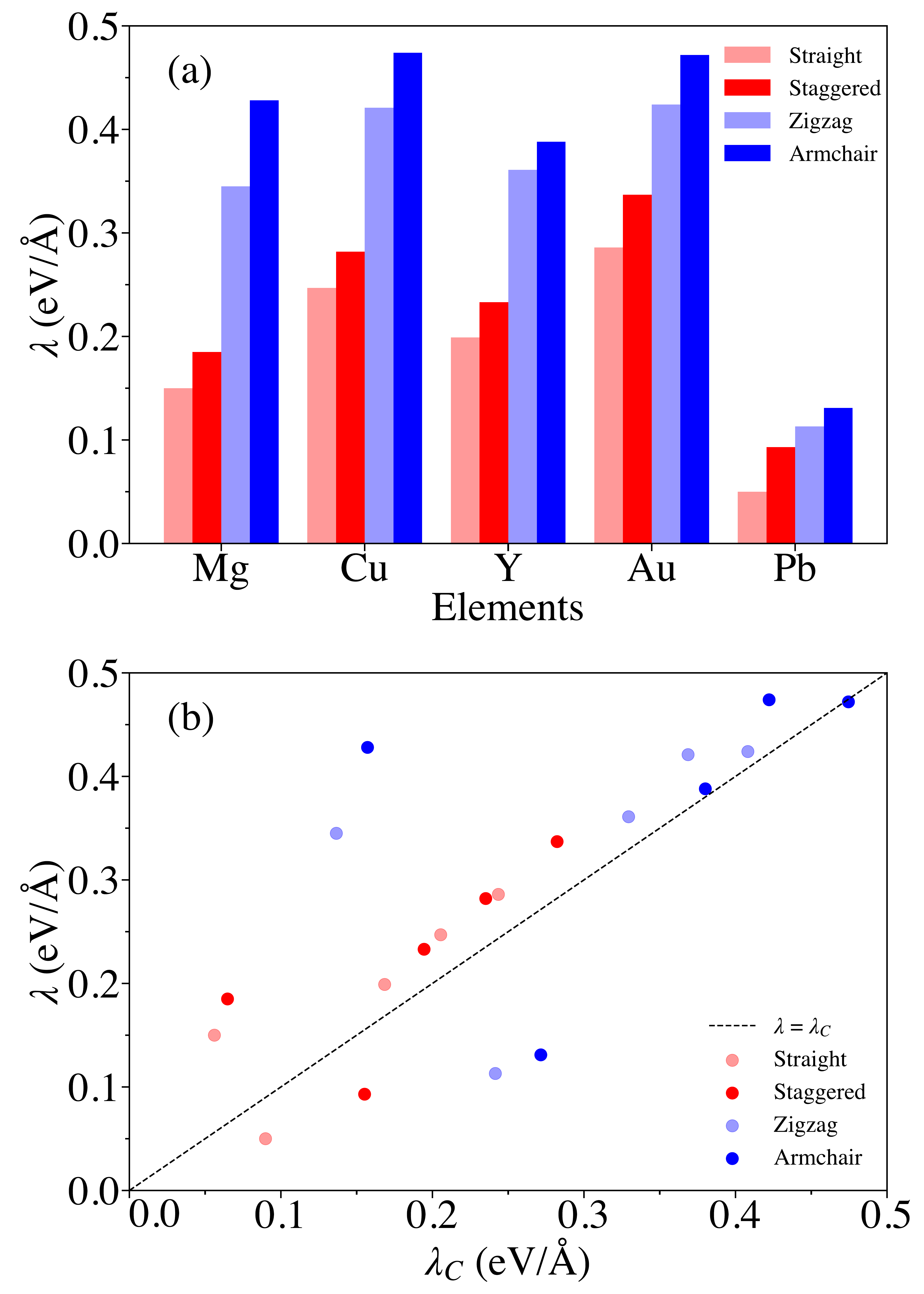}
    \caption{Edge energetics. a) Energies for all edges and elements. b) The comparison between calculated edge energies ($\lambda$) and estimate based on coordination numbers ($\lambda_C$), Eq.~(\ref{eq:gamma_model}).}
    \label{fig:edge_energies}
\end{figure}

The edge energies are around $0.05-0.5$ eV/\AA\ (Fig.~\ref{fig:edge_energies}a). They are smaller for Pb and larger for Cu and Au, following the surface energies of these metals \cite{LEE2SurfaceEnegries2018}. As shown previously, edge energies are strongly correlated with cohesion energy \cite{nevalaita2018atlas, nevalaita2018beyond}, and they depend on the Wigner-Seitz radius the same way as do surface energies \cite{nevalaita2018beyond}. The edge energies of metallenes are roughly half as small compared to those of 2D covalent materials like graphene ($1-1.5$~eV/\AA) and \ce{MoS2} ($\lesssim 1$~eV/\AA) \cite{koskinen2008self,EdgesBeyondpekka2009, DFTMoS2Edges2013}. 

Among the different edge types and due to the larger density of undercoordinated edge atoms, hexagonal edges have smaller edge energies than buckled honeycomb edges. The straight edges are the cheapest, the armchair edges are the most expensive. The edge energy of Pb(sta) is $86$~\%\ larger than Pb(str), demonstrating an unexpectedly large dependence on edge orientation. For other elements, the relative difference between straight and staggered edges is smaller. For the buckled honeycomb lattices, the zigzag edges are energetically more favorable than the armchair edges. These results are consistent with similar trends in other 2D covalent materials like graphene, \ce{MoS2}, and hexagonal boron nitride \cite{koskinen2008self, EdgesBeyondpekka2009, liu2022recent}. The relative difference between zigzag and armchair edges is the highest for Mg ($25$~\%) and the lowest for Y ($7$~\%). Similarly, the relative difference between hexagonal and buckled honeycomb edges is the highest for Mg ($57$~\%\ on average) and the lowest for Au ($30$~\%\ on average).

A simple non-linear bond-cutting model based on coordination number $C$ explains well the edge energy trend among edge types and lattices \cite{bondCutting1992calculated, CohCoord2017}. While a linear bond-cutting model (cohesion $\propto C$) fails to describe the edge energies just as it fails to describe metals' surface energies, a non-linear model (cohesion $\propto $ $\sqrt{C}$) is more reasonable (Fig.~\ref{fig:edge_energies}b). This model approximates the edge energy as
\begin{equation}
    \lambda_C =|\varepsilon_{2D}| \cdot \frac{\sqrt{C_0}- \sqrt {C_m}}{\sqrt{C_0}}\eta,
    \label{eq:gamma_model}
\end{equation}
where $C_0$ is the \mb{coordination number obtained from 2D bulk symmetry}, $C_m$ is the mean coordination number of edge atoms (undercoordinated atoms at the edge), and $\eta$ is the density of those atoms per unit length \cite{Bond-cuttingmclachlan1957surface, bondCutting1992calculated}. \mb{Only $\varepsilon_{2D}$ is obtained from a separate DFT calculation; all other parameters are given directly by the structure.} A simple bond-counting exercise yields $\lambda_C=|\varepsilon_{2D,hex}|\cdot 0.183/a$ for straight, 
$\lambda_C=|\varepsilon_{2D,hex}|\cdot 0.212/a$ for staggered, $\lambda_C=|\varepsilon_{2D,bhc}|\cdot 0.300/a$ for zigzag, and 
$\lambda_C=|\varepsilon_{2D,bhc}|\cdot 0.347/a$ for armchair edges, where $a$ is the lattice constant (\emph{cf.} Fig.~\ref{fig:systems}a). 

Apart from the outliers of Pb and Mg, the main trend for the bond-cutting model is to slightly underestimate the DFT value for the edge energy. \mb{The outliers arise due to the choice of the nonlinear bond-cutting model and not necessarily due to a particularly peculiar behavior of the elements. Further refinement of the model, such as refining the exponent (e.g., fitting it as suggested in Ref. \citenum{abidi2022optimizing}), could potentially eliminate these outliers. However, such refinements would unnecessarily complicate the model and reduce its transparency.}
The model of Equation (\ref{eq:gamma_model}) works best for straight edge, the simplest edge type. The non-linear bond-cutting model is particularly consistent in reproducing the relative differences between straight and staggered edge energies.

% the bond-cutting model overestimates...

\subsection{Edge relaxations and stresses}
The edge relaxation depends sensitively on both element and edge type. Atoms move outward in straight edges and inward in staggered edges, except for Mg. Consequently, compared to an ideal triangular lattice, bond angles increase on the straight edge and decrease on the staggered edge. No hexagonal edge has out-of-plane movement. In buckled honeycomb lattices, the relaxation is more complex and element-specific. For zigzag edges, the atoms relax outward for Au and Cu and inward for other elements. \mb{The different behavior can be attributed to Au and Cu’s fully occupied d-bands far below the Fermi level. In Au and Cu, no partially filled d-states are available for directional rehybridization at undercoordinated straight and zigzag edges (which are less undercoordinated for the outermost atoms, compared to sta and ac). Consequently, dominated by broad s- and p-type bands, the conduction electrons drive outward relaxation that helps lower electron repulsion at the edge. In contrast, in Mg, Y, and Pb, the partially occupied valence orbitals (d-electrons for Y, p-electrons for Pb, and s- and p-electrons for Mg) may compensate for lost coordination and strengthen orbital overlap by inward relaxation.} 

\begin{figure}
    \centering
    \includegraphics[width=\linewidth]{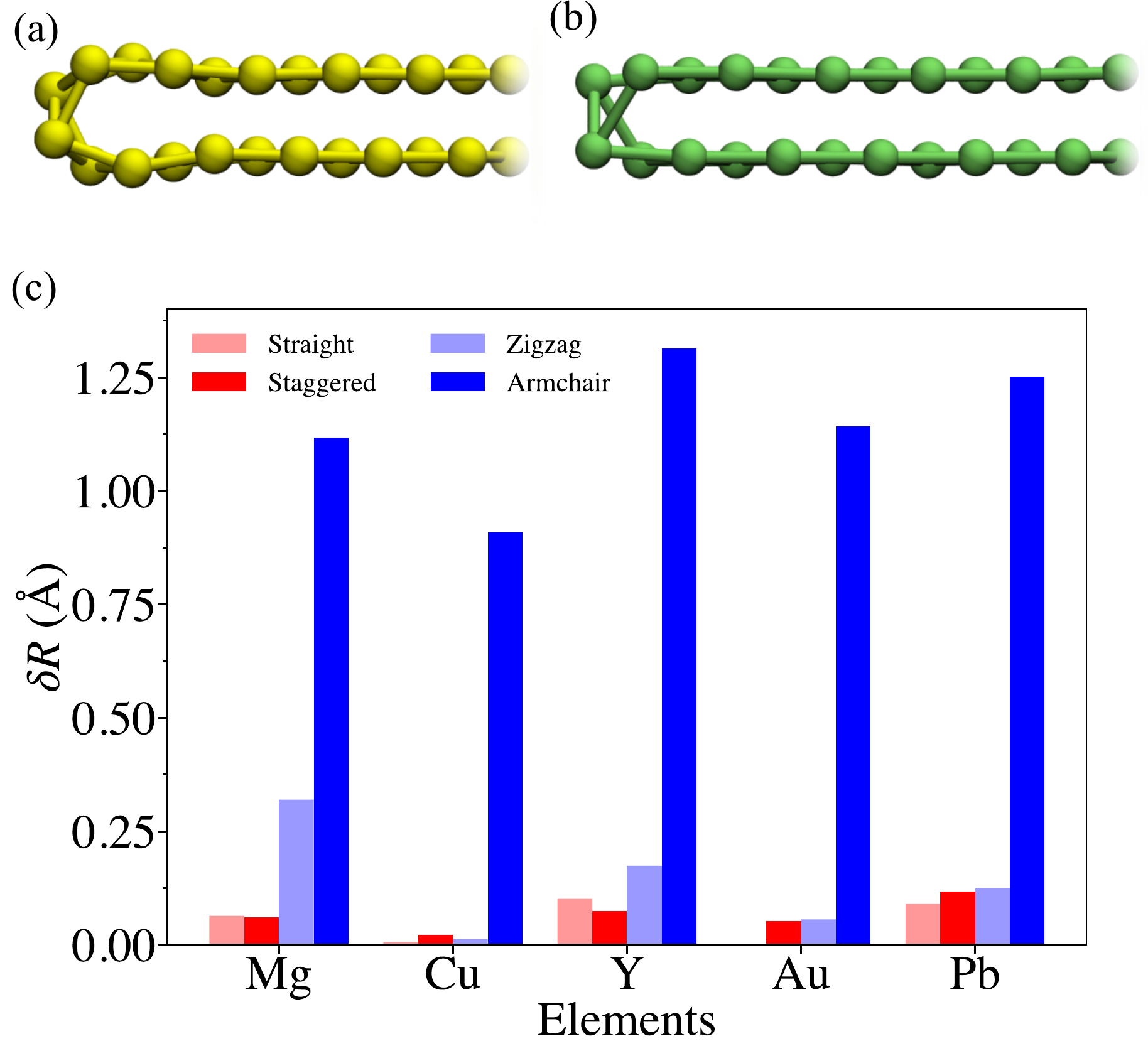}
    \caption{Relaxation of the metallene edges. a) Closed loop geometry of the Au(ac) edge. b) The geometry of the Mg(ac) edge. c) The magnitude of edge relaxation in terms of root-mean-square bond length variations at the edge [Eq.~(\ref{eq:delta_R})].}
    \label{fig:relaxation}
\end{figure}

For armchair edges, the atoms relax outward for Pb and Cu and inward for other elements. The buckling thickness reduces for both edges and all elements. A fascinating exception is Au(ac), which spontaneously relaxes into a closed edge, forming a (giant) single-wall goldene nanotube (Fig. \ref{fig:relaxation}a) \cite{Liu2020, Bi_2008, C3NR33658A}.
A similar phenomenon is familiar from graphene bilayer edges, where a loop forms between the two layers to eliminate dangling bonds and increase edge stability \cite{GrapheneBilayerEdges2009}. It is curious that a metallic element like Au should behave similarly to a covalent material in this respect.

To quantify the magnitude of the edge relaxation, we used the equation 
\begin{equation}
    \delta R = \sqrt{ \sum_{i \in \text{edge}} \sum_{j\; \text{n.n.}} (R_{ij} - R_{0})^2/N_\text{bonds}},
    \label{eq:delta_R}
\end{equation}
where $N_{bonds}$ is number of nearest-neighbor bonds at the edge, $R_{0}$ is the 2D bulk bond length, $R_{ij}$ is the distance between edge atom $i$ and its nearest neighbor $j$. The most striking result is the prominent relaxation of the armchair edges for all elements. The zigzag edges relax much less. Among hexagonal edges, the staggered edges relax more than straight edges, except for Mg and Y. The relaxation at the Au and Cu straight edges is virtually nonexistent.

\begin{figure}[t!]
    \centering
    \includegraphics[width=1.0\linewidth]{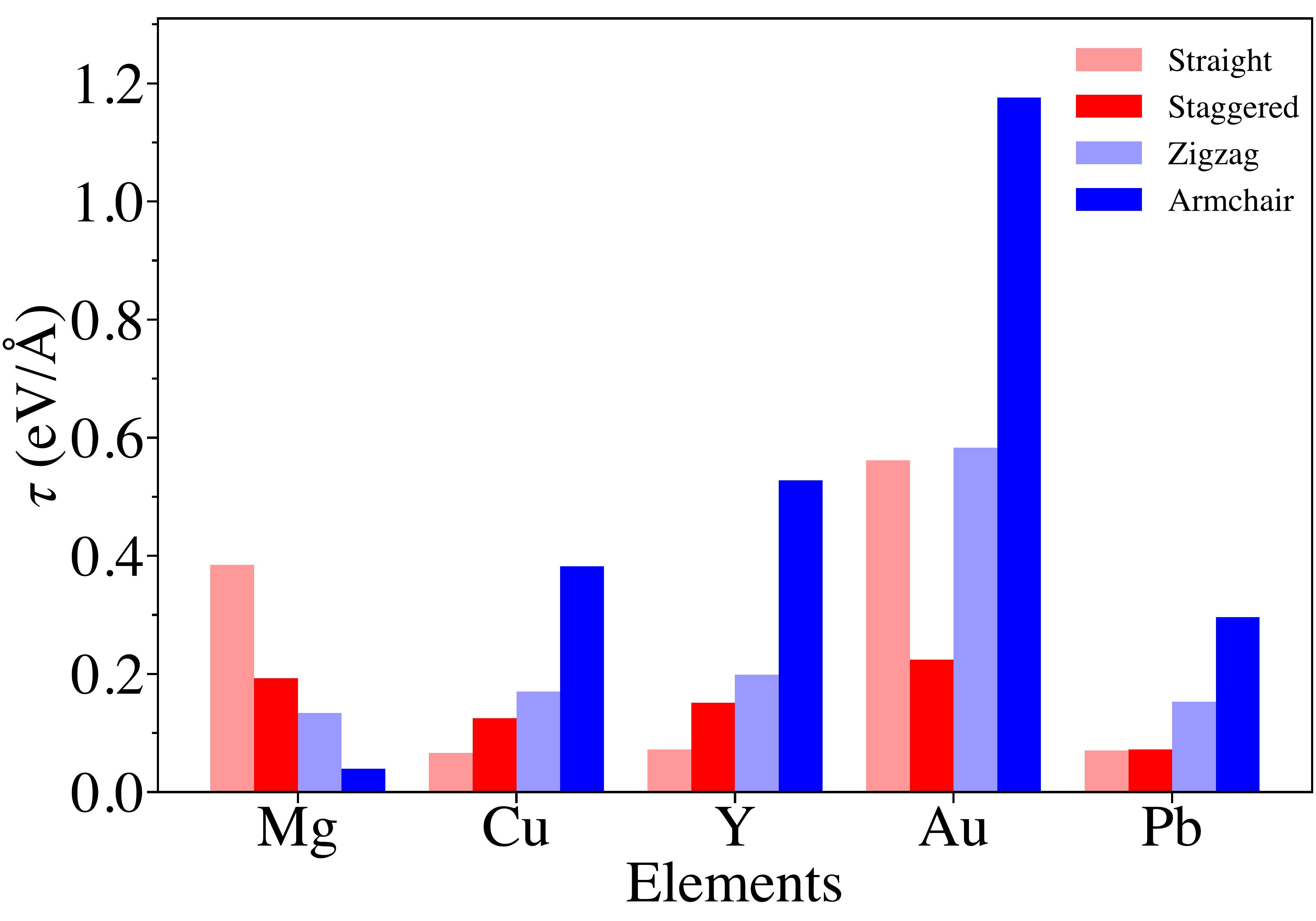}
    \caption{Edge stresses for the studied systems. All stresses are tensile.}
    \label{fig:edge_stresses}
\end{figure}

Like in \ce{MoS2} but unlike in graphene, the stresses are tensile ($\tau>0$) for all edges, meaning a decreasing energy for an edge shrinking along its length (Fig.~\ref{fig:edge_stresses}) \cite{DFTMoS2Edges2013}. Stresses are between $0.04-0.55$~eV/\AA, smaller than in graphene \cite{huang2009quantum} but following a similar trend in the edge energies (which have the same units as edge stresses). Again, an exception is Au(ac), which has a sizable stress of $1.15$~eV/\AA. The stress arises from forming the closed edge, which requires the formation of new bonds parallel to the edge, causing strained bond lengths and angles responsible for the stress.

The stresses for Cu, Y, and Pb behave the same way as edge energies, except for Mg where this trend reverses completely. Au again is a special case where $\tau$ follows the trend of $\lambda$ for the buckled honeycomb lattice but reverses for the hexagonal lattice. The edge energies and stresses have a distinct connection to edge reconstruction. As discussed in Ref.~\cite{LEE2SurfaceEnegries2018}, the difference between surface tension and surface energy drives surface reconstruction. Analogously, here the difference between edge stress and edge energy can drive edge reconstruction. The larger the difference $\tau-\lambda$, the greater the edge relaxation (Fig.~\ref{fig:relaxation}b).
The driving force is by far the largest for Au(ac) edge, the curious case of the closed edge. The difference $\tau-\lambda$ is negative for edges where atoms relax inward.

\subsection{Electronic Structure}

We also investigated the electronic structures at metallene edges. For this purpose, we probed the variation of the local density of states (DOS) as a function of distance from the edge. The probe was constructed by calculating the average local density of states using the expression 
\begin{equation}
DOS_F(i) = \frac{\int_{-\infty}^{+\infty} g(\varepsilon) \text{DOS}_i(\varepsilon) \, d\varepsilon}{\int_{-\infty}^{+\infty} g(\varepsilon)d\varepsilon}   . 
\end{equation}
Here $\varepsilon$ is the single-particle energy, $\text{DOS}_i$ is the density of states projected on atom $i$, and $g(\varepsilon)=\exp[-(\varepsilon - \varepsilon_{F})^2/2\sigma^2]$ is a Gaussian envelope function with $\sigma = 0.2$ eV and Fermi level $\varepsilon_{F}$ \cite{abidi2022optimizing}. Thus, $DOS_F(i)$ represents an average density of states at atom $i$ near the Fermi-level. 

\begin{figure}[ht!]
    \centering
    \includegraphics[width=\linewidth]{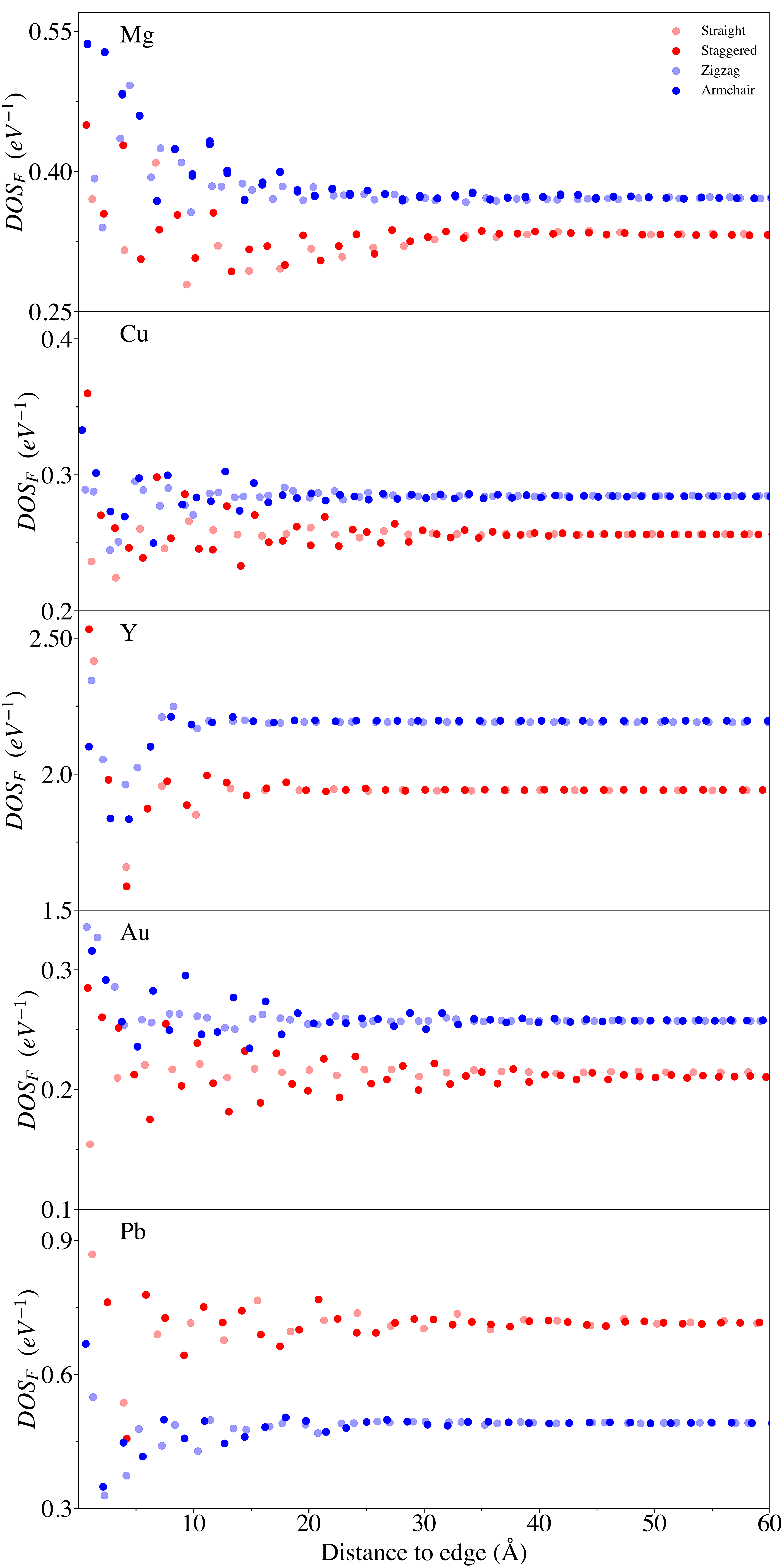}
    \caption{Atom-projected densities of states for all elements and edge types.}
    \label{fig:pdos}
\end{figure}

The values of $DOS_F(i)$ at the edge differ significantly from bulk values (Fig.~\ref{fig:pdos}; see ESI for detailed plots of $DOS_i$). While the differences arise due to the lower coordination of edge atoms, they penetrate nanometers deep into the bulk, far beyond the mere edge atoms. For Mg, Cu, and Au, $DOS_F(i)$ converges towards the bulk within the length $\sim 3$~nm for all edges, although the $DOS_F(i)$ variation at the edges depends on edge type. The similar convergence of $DOS_F(i)$ across different edge types suggests that, at a large scale, these element's electronic structures are relatively insensitive to the specifics of edge geometry. This behavior can be attributed to these elements' filled or nearly filled $s$- and $d$-orbitals, which promotes delocalized metallic bonding. For Y, the convergence depends on the lattice in question. The $DOS_F(i)$ variation in buckled honeycomb edges is twice as large as in hexagonal edges. Such edge dependence presumably arises from the partially filled $4d$-orbital, which makes Y sensitive to changes in coordination and edge symmetry. For Pb, the edge dependency is even more pronounced. The staggered edge has the longest ($\sim 5$~nm) and zigzag the shortest ($\sim 2$~nm) convergence length. These differences demonstrate the presence of edge-specific states or dangling bonds and an involved connection between $p$-orbitals and the edge geometry. 

\mb{The mechanism that makes Mg, Cu, and Au different from Y and Pb in edge sensitivity rests on orbital occupations of frontier orbitals and the resultant directional bonding of undercoordinated atoms \cite{CohCoord2017}. In Mg, Cu, and Au, states near Fermi-level derive from deep d- or (nearly) filled s-bands, which are relatively insensitive to local changes in coordination. Therefore, edge-specificities in DOS are mild and convergence lengths similar. In contrast, Y’s partially filled 4d-orbitals and Pb’s partially occupied p-orbitals are more sensitive to coordination and bond angle changes. These sensitivities cause more pronounced variations in the local DOS. For Y and Pb, the directional nature of partially filled d- and p-orbitals makes small changes in edge geometry to split or shift these states \cite{dOrbital}, forming states akin to “dangling bonds” that depend on edge type. The large orbital reconfigurations available in partially filled bands are sufficient to yield a greater edge type -dependence for Y and Pb than for Mg, Cu, or Au. Spin–orbit coupling would modify Pb’s electronic structure even further, as reported for surfaces \cite{Klaus1994}.} The metallene edge states still need further investigation for topological protection or other potentially useful electronic properties.

%Although spin–orbit coupling further modifies Pb’s electronic structure in real material as in the case of surfaces \cite{Klaus1994}, even without it, the large orbital reconfigurations available to partially filled p-states are sufficient to yield a much greater edge-geometry dependence than seen in Mg, Cu, or Au

\vspace{5mm}
\section{Summary and conclusions}
In summary, we have investigated the basic properties of edges in atomically thin metallenes made of five representative elements from different parts of the periodic table. We focused on hexagonal and buckled honeycomb lattices with two edge orientations and four edge types for each metallene. 

The edge energies for all metallenes followed the same trend, which could be explained reasonably well by a non-linear bond-cutting model, analogously to surface energies of the same elements \cite{LEE2SurfaceEnegries2018, bondCutting1992calculated, CohCoord2017}. The edge stresses in all metallenes were tensile, which suggests the edges' tendency to shrink the edges along their length. The tensile stress improves the mechanical stability of the edges compared to compressive stress, which---conversely---tends to create out-of-plane rippling in 2D materials \cite{EdgeStressInducedSpontaneous20212, shenoy2008edge}. Still, the slow convergence of the local density of states while moving from the edge into the 2D bulk suggests that edge stresses (and edge energies alike) may originate not only from the undercoordinated edge atoms but from deeper within the bulk. Therefore, the pairwise atomic interactions and local stresses near the edges should be investigated in more detail. The armchair edge of Au revealed the most peculiar behavior by relaxing into a closed loop, forming a partially squashed Au nanotube; this behavior also deserves further investigation. 

To conclude, this work presents multifaceted new knowledge about the edges of atomically thin metallenes. This knowledge, which still requires further completion, is crucial for our experimental attempts to synthesize more stable and versatile lateral interfaces between metallenes and covalent templates.

\begin{acknowledgments}
We acknowledge the Vilho, Yrjö, and Kalle Väisälä Foundation of the Finnish Academy of Science and Letters and the Jane and Aatos Erkko Foundation for funding (project EcoMet) and the Finnish Grid and Cloud Infrastructure (FGCI) and CSC—IT Center for Science for computational resources.
\end{acknowledgments}

%\bibliography{refs}

\end{document}